\documentclass[twoside,twocolumn,english,aps,pra,reprint,amsmath,amssymb]{revtex4-2}
\usepackage[T1]{fontenc}
\usepackage[utf8]{inputenc}
\usepackage{verbatim}
\usepackage{mathtools}
\usepackage{amsmath}
\usepackage{amssymb}
\usepackage{cancel}
\usepackage{mathdots}
\usepackage{stmaryrd}
\usepackage{stackrel}
\usepackage{graphicx}
\PassOptionsToPackage{version=3}{mhchem}
\usepackage{mhchem}
\usepackage{esint}

\makeatletter
\usepackage{paralist}

\usepackage[colorlinks=true,
            linkcolor=red,      
            urlcolor=magenta,   
            citecolor=blue,     
            anchorcolor=blue]{hyperref}

\begingroup
\footnotetext{These authors contributed equally to this work.}

\usepackage{mciteplus}

\usepackage{microtype} 

\usepackage{newtxtext,newtxmath}  

\usepackage{booktabs}   
\usepackage{siunitx}    

\makeatother

\usepackage{babel}
\begin{document}
\selectlanguage{english}
\title{Reshaping nonclassical properties and metrological performance of entangled coherent states via post-selected von Neumann measurements}
\author{Janarbek Yuanbek$^{1,2,4}$\textsuperscript{\textsection}, Bruno
Tenorio$^{3}$\textsuperscript{\textsection}}
\author{Yusuf Turek$^{4}$}
\email{Corresponding author: yusufu1984@hotmail.com}

\affiliation{$^{1}$\textup{State Key Laboratory of Semiconductor Physics and Chip Technologies, Institute of Semiconductors, Chinese Academy of Sciences, Beijing 100083, China}}
\affiliation{$^{2}$\textup{Center of Materials Science and Opto-Electronic Technology, University of Chinese Academy of Sciences, Beijing 100049, China}}
\affiliation{$^{3}$\textup{Wolfram Research South America, Lima 15076, Peru}}
\affiliation{$^{4}$\textup{School of Physics, Liaoning University, Shenyang, Liaoning 110036, China}}
\date{\today}
\begin{abstract}
In quantum metrology, measurements are usually treated as passive
readout processes. Here we investigate whether post-selected von Neumann
measurements (PVNMs) can be used as an active resource to reshape
the nonclassical properties of a two-mode entangled coherent state
(ECS). By analyzing the finite-coupling post-selected state, we show
that PVNMs can enhance quadrature squeezing and sum squeezing, increase
the Wigner-function negativity, and strengthen bipartite correlations,
as witnessed by the Hillery--Zubairy criterion and linear entropy.
We further evaluate the quantum Fisher information and the corresponding
quantum Cramér--Rao bound for phase estimation, and discuss the trade-off
between metrological gain and measurement-induced disturbance through
the fidelity. Our scheme exhibits a phase-sensitivity advantage over
standard ECS metrology for large average photon numbers. Our results
suggest that PVNMs provide a tunable route for engineering nonclassical
resources in continuous-variable sensing protocols.
\end{abstract}
\maketitle

\section{Introduction}

Quantum metrology \citep{PhysRevLett.96.010401} has become a central
topic in quantum optics because it offers the possibility of parameter
estimation beyond classical limits, %
{} by exploiting nonclassical resources such as entanglement and interference
\citep{PhysRevLett.130.170801}. A variety of probe states have been
proposed for this purpose, including NOON states \citep{PhysRevLett.94.090502,PhysRevLett.98.223601},
“bat” states \citep{PhysRevA.81.043624}, and entangled coherent states
(ECS) \citep{PhysRevLett.107.083601,PhysRevA.45.6811,jeon2024experimental}.
Among these, ECS have been shown to offer competitive phase-estimation
sensitivity compared with NOON states, “bat” states, and uncorrelated
states at the same average photon number, while also exhibiting favorable
robustness to losses in certain regimes. These advantages are commonly
quantified using the quantum Fisher information, which provides a
fundamental bound on the achievable precision in parameter estimation.

Despite these advances, existing analyses are predominantly based
on standard measurement paradigms, where the role of measurement is
treated as passive and does not fundamentally alter the structure
of the probe state \citep{PhysRevD.40.2112,PhysRevA.41.11}. This
assumption may become inadequate when considering weak measurement
protocols, in which measurement itself can actively reshape the statistical
and coherence properties of quantum systems. Within this framework,
pre-selection, post-selection, and weak system--pointer coupling
jointly enable information extraction with limited disturbance, giving
rise to weak values that may lie outside the eigenvalue spectrum and
can be used for signal amplification in appropriate regimes \citep{PhysRevLett.60.1351,DELIMABERNARDO2014194,janarbek,PhysRevX.4.011032}.
Such features have been successfully exploited in single-mode settings
\citep{PhysRevA.81.033813} and have been suggested to influence quantum
statistical properties \citep{qmen-RevModPhys.81.865,qmen-Vedral2014QuantumE,qmen-Tavares2023QuantumE}
more generally, indicating that weak measurements may offer capabilities
beyond those captured by conventional metrological approaches.

However, how these measurement-induced effects extend to entangled
states remains less explored. In particular, the non-destructive manipulation
of nonclassical features in ECS poses a significant challenge for
practical implementations. The presence of post-selection can substantially
modify measurement statistics and quantum correlations, potentially
leading to regimes where precision limits deviate from those predicted
by standard quantum Fisher information analysis \citep{jy6g-jp7n,PhysRevLett.115.120401,PhysRevA.106.022619}.
Understanding this interplay is therefore essential for determining
whether weak measurement can enhance, preserve, or even degrade the
metrological advantages of ECS. Motivated by these considerations,
it is crucial to establish a systematic framework for analyzing post-selected
von Neumann measurements (PVNMs) \citep{vonNeumann+2018} acting on
ECS. Such an investigation not only clarifies the role of measurement
in quantum metrology but also opens the possibility of leveraging
weak measurement as an active resource for optimizing quantum sensing
protocols beyond conventional paradigms.

Unlike previous weak-value amplification studies that mainly focus
on anomalous pointer shifts and parameter estimation, the present
work investigates how post-selected von Neumann measurements reshape
the nonclassical properties of entangled coherent states. In this
work, we study PVNMs applied to a two-mode ECS and analyze their influence
on squeezing, Wigner-function negativity, entanglement, and phase-estimation
performance. By tuning the system--pointer coupling strengths, we
show that PVNMs can modify the quantum state in a controllable manner,
which in certain parameter regimes leads to enhanced quadrature and
sum squeezing, increased Wigner negativity, and stronger bipartite
entanglement as quantified by the Hillery--Zubairy criterion and
linear entropy. We further examine the quantum Fisher information
and the corresponding quantum Cramér--Rao bound, showing that the
measurement-induced modification can improve phase sensitivity within
the explored parameter regime. Compared with standard ECS metrology
\citep{PhysRevLett.107.083601}, our post-selected scheme achieves
a distinct phase-sensitivity advantage for large average photon numbers.
At the same time, fidelity analysis reveals the expected trade-off
between resource enhancement and state disturbance.

The paper is organized as follows. In Sec. \ref{2}, we construct
the theoretical model using the PVNMs framework and derive the final
pointer state. Sec. \ref{3}, we employ this model to analyze the
effects of PVNMs on the quadrature and sum squeezing properties of
the entangled coherent state. In Sec. \ref{4}, we examine the measurement
back-action in phase space by studying the evolution of the joint
and single-mode Wigner functions. In Sec. \ref{5}, we quantify the
modification of two-mode entanglement using the Hillery-Zubairy criterion
and linear entropy. In Sec. \ref{6}, we assess the metrological utility
of the enhanced state by calculating the quantum Fisher information.
In Sec. \ref{7}, we analyze the fidelity between the initial and
post-selected pointer states. In Sec. \ref{8}, we briefly discuss
the influence of open system effects on the dynamics. Finally, in
Sec. \ref{9}, we summarize the key findings and provide an outlook
for future research. Throughout the paper, we set $\hbar=1$.

\section{Theoretical Framework\label{2}}

\begin{figure}
\centering
\begin{centering}
\includegraphics[scale=0.07]{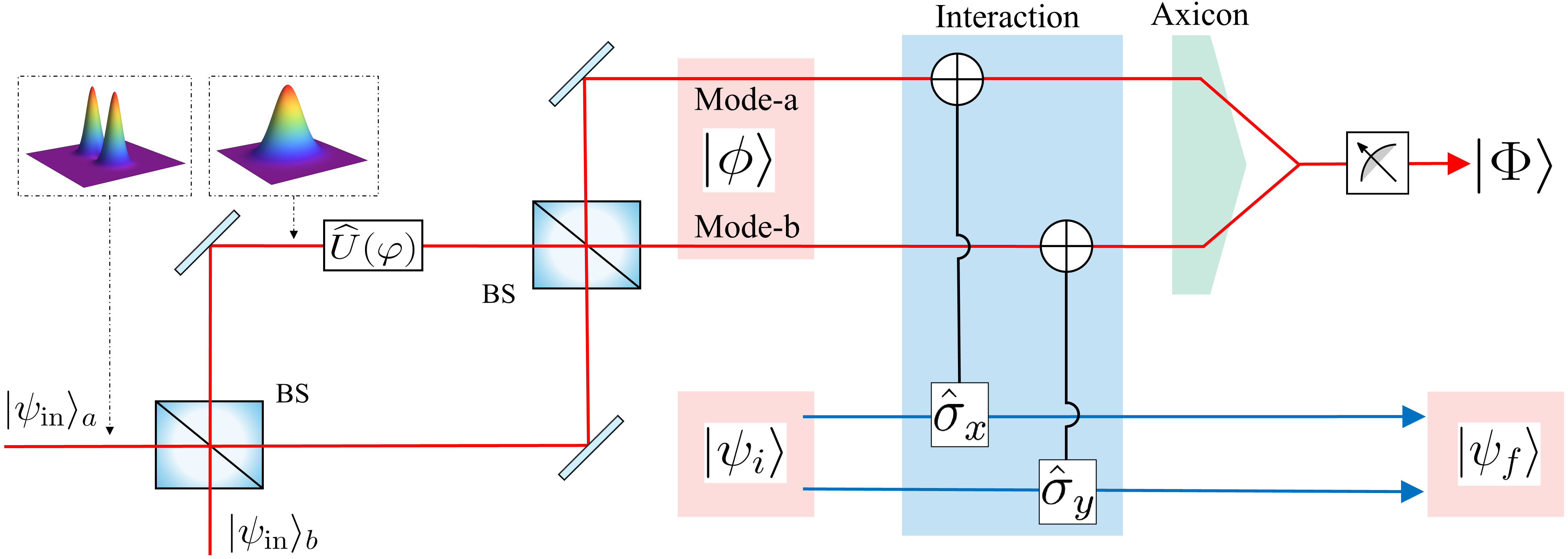} 
\par\end{centering}
\caption{Schematic illustration of ECS preparation and PVNMs. An ECS is generated
by interfering coherent inputs at a 50:50 beam splitter and applying
the phase shift $\hat{U}(\varphi)$. The system is then pre-selected
in $|\psi_{i}\rangle$, interacts with the pointer, and is post-selected
in $|\psi_{f}\rangle$, yielding the final pointer state $|\Phi\rangle$.\label{fig:mode}}
\end{figure}

ECS are nonclassical states that arise from the superposition of coherent
light fields across various spatial modes, combining the experimental
accessibility of coherent states with intrinsically quantum entanglement
properties \citep{PhysRevA.86.043828}. As illustrated in Fig. \ref{fig:mode},
an ECS can be generated by injecting an even cat state $\vert\psi_{\mathrm{in}}\rangle_{a}=\mathcal{N}(|\alpha/\sqrt{2}\rangle+|-\alpha/\sqrt{2}\rangle)$
\citep{PhysRevA.110.L010602}, where $\mathcal{N}=[2(1+e^{-2|\alpha|^{2}})]^{-1/2}$
is the normalization factor \citep{Int} and a coherent state $\vert\psi_{\mathrm{in}}\rangle_{b}=|\alpha/\sqrt{2}\rangle$
into the two input ports of a 50:50 beam splitter (BS) \citep{PhysRevA.80.022111,PhysRevLett.107.083601}.
After the unitary transformation implemented by the beam splitter,
is given by $\vert\phi_{\mathrm{in}}\rangle=\vert\psi_{\mathrm{in}}\rangle_{a}\otimes\vert\psi_{\mathrm{in}}\rangle_{b}=\mathcal{N}(|\alpha\rangle_{a}\vert0\rangle_{b}+|0\rangle_{a}\vert\alpha\rangle_{b}).$
Here, $\alpha=re^{i\mu}$, and $|0\rangle$ and $|\alpha\rangle$denote
the vacuum and coherent states in spatial modes $a$ and $b$, respectively%
. To study phase sensing, we apply a phase shift to mode $b$ through
$\hat{U}(\varphi)=\mathbb{I}\otimes e^{i\varphi\hat{n}_{b}}$ , where
$\hat{n}_{b}=\hat{b}^{\dagger}\hat{b}$ is the photon-number operator
of mode $b$, and $\mathbb{I}$ denotes the identity operator acting
on mode $a$. The action of this operation on a coherent state is
given by $e^{i\varphi\hat{n}_{b}}|\alpha\rangle_{b}=|\alpha e^{i\varphi}\rangle_{b}$.
Finally, the output ECS is obtained as

\begin{equation}
|\phi\rangle=\hat{U}(\varphi)|\phi_{\mathrm{in}}\rangle=\mathcal{N}(|\alpha\rangle_{a}|0\rangle_{b}+|0\rangle_{a}|\alpha e^{i\varphi}\rangle_{b}).\label{eq:pointer-state}
\end{equation}

For simplicity, %
this work analyzes the impact of PVNMs on the properties of ECS and
their applications in quantum precision measurement, where the total
system pointer Hamiltonian is decomposed into three fundamental terms
governing distinct physical processes during the interaction. The
total Hamiltonian is written as

\begin{equation}
\hat{H}=\hat{H}_{\mathrm{s}}+\hat{H}_{\mathrm{p}}+\hat{H}_{\mathrm{int}},
\end{equation}
where $\hat{H}_{\mathrm{s}}$ and $\hat{H}_{\mathrm{p}}$ describe
the system and pointer, respectively, and $\hat{H}_{\mathrm{int}}$
describes their interaction. Within ideal measurement theory, the
choice of $\hat{H}_{\mathrm{s}}$ and $\hat{H}_{\mathrm{p}}$ is inconsequential
to the outcomes; only $\hat{H}_{\mathrm{int}}$ determines the measurement
strength and post-selection statistics. Without loss of generality,
the interaction Hamiltonian is chosen as \citep{PhysRevLett.109.040401}

\begin{equation}
\hat{H}_{\mathrm{int}}=g_{a}\hat{\sigma}_{x}\otimes\hat{P}_{x}+g_{b}\hat{\sigma}_{y}\otimes\hat{P}_{y},\label{eq:introduction hamli}
\end{equation}
where $g_{a}$ and $g_{b}$ are the system--pointer coupling strengths.
The measured observables $\hat{\sigma}_{x}$ and $\hat{\sigma}_{y}$
are expressed in their respective eigenbases as 

\begin{align}
\hat{\sigma}_{x} & =|H\rangle\langle V|+|V\rangle\langle H|,\\
\hat{\sigma}_{y} & =i|V\rangle\langle H|-i|H\rangle\langle V|,
\end{align}
where $|H\rangle$ and $|V\rangle$ denote the horizontal and vertical
polarization states. The pointer momenta$\hat{P}_{x}$ and $\hat{P}_{y}$are
conjugate to the pointer quadratures $\hat{X}$ and $\hat{Y}$, satisfying
$[\hat{X},\hat{P}_{x}]=[\hat{Y},\hat{P}_{y}]=i$. Both operators can
be expressed via annihilation ($\hat{a}$, $\hat{b}$) and creation
($\hat{a}^{\dagger}$, $\hat{b}^{\dagger}$) operators as $\hat{P}_{x}=i(\hat{a}^{\dagger}-\hat{a})/2\sigma$
and $\hat{P}_{y}=i(\hat{b}^{\dagger}-\hat{b})/2\sigma$ \citep{PhysRevA.41.1526},
where $\sigma=\sqrt{1/2m\omega}$ is the ground-state width of the
pointer, determined by its mass $m$ and oscillation frequency $\mathrm{\omega}$.
As shown in Fig. \ref{fig:mode}, we illustrate the experimental setup
of weak interactions. Based on this, the initial composite state is
given by

\begin{equation}
\vert\Psi_{in}\rangle=\vert\psi_{i}\rangle\otimes\vert\phi\rangle.\label{int}
\end{equation}
Here the initial system state is prepared as $\vert\psi_{i}\rangle=\cos(\theta/2)\vert H\rangle+e^{i\delta}\sin(\theta/2)\vert V\rangle$,
where $\delta\in[0,2\pi]$ and $\theta\in[0,\pi)$. In a PVNMs, the
system is defined by both pre- and post-selection of its state. We
prepare the system in an initial state $\vert\psi_{i}\rangle$ and
then we couple it to the pointer state. After some interaction time
$t$, we post-select a system state $\vert\psi_{f}\rangle$, and obtain
information on the physical quantities $\hat{\sigma}_{x}$ and $\hat{\sigma}_{y}$
from the pointer wave function. For definiteness, we set the post-selected
state to $\vert\psi_{f}\rangle=\vert H\rangle$. The interaction acts
for a finite duration $t$, so the unitary evolution generated by
$H_{int}$ is 
\begin{equation}
\hat{U}(t)=e^{-i\int^{t}_{0}H_{int}\,dt'}=e^{-itg_{a}\hat{\sigma}_{x}\otimes\hat{P}_{x}-ig_{b}t\hat{\sigma}_{y}\otimes\hat{P}_{y}},
\end{equation}
Since the operators $\hat{\sigma}_{x}$ and $\hat{\sigma}_{y}$ satisfy
$\hat{\sigma}^{2}_{x}=\hat{\sigma}^{2}_{y}=\mathbb{I}$, the evolution
operators can be written as:
\begin{align}
e^{-ig_{a}t\hat{\sigma}_{x}\otimes\hat{P}_{x}} & =\frac{1}{2}\left[\Upsilon_{+}\otimes D_{1}[\frac{s_{1}}{2}]+\Upsilon_{-}\otimes D^{\dagger}_{1}[\frac{s_{1}}{2}]\right],\\
e^{-ig_{b}t\hat{\sigma}_{y}\otimes\hat{P}_{y}} & =\frac{1}{2}\left[\Upsilon^{\prime}_{+}\otimes D_{2}[\frac{s_{2}}{2}]+\Upsilon^{\prime}_{-}\otimes D^{\dagger}_{2}[\frac{s_{2}}{2}]\right],
\end{align}
where $\Upsilon_{\pm}=\mathbb{I}\pm\hat{\sigma}_{x}$ and $\Upsilon'_{\pm}=\mathbb{I}\pm\hat{\sigma}_{y}$.
The displacement operator is defined by $D(\alpha)=\exp\bigl[\alpha\hat{a}^{\dagger}-\alpha^{\ast}\hat{a}\bigr]$,
and the coupling strength parameter is $s_{1}=g_{a}t/\sigma$, and
$s_{2}=g_{b}t/\sigma$. For real $s_{1,2}$ this yields $D_{1}[s_{1}/2]=\exp[s_{1}(\hat{a}^{\dagger}-\hat{a})/2]$
and $D_{2}[s_{2}/2]=\exp[s_{2}(\hat{b}^{\dagger}-\hat{b})/2]$. After
the interaction, the initial state becomes 
\begin{equation}
\left|\Psi\right\rangle =\hat{U}(t)\left|\Psi_{in}\right\rangle .
\end{equation}

In PVNMs, the state $\left|\Psi\right\rangle $ is projected onto
the post-selected state $\vert\psi_{f}\rangle$, yielding an unnormalized
pointer state $\vert\tilde{\Phi}\rangle=\langle\psi_{f}\vert\Psi\rangle$.
The final normalized probe state is given by 
\begin{align}
\left|\Phi\right\rangle  & =\frac{\vert\tilde{\Phi}\rangle}{\sqrt{P_{s}}}\nonumber \\
 & =\frac{\kappa}{4}\Bigg\{ A_{+}D_{1}[\frac{s_{1}}{2}]D_{2}[\frac{s_{2}}{2}]+A_{-}D^{\dagger}_{1}[\frac{s_{1}}{2}]D^{\dagger}_{2}[\frac{s_{2}}{2}]\nonumber \\
 & +B_{+}D^{\dagger}_{1}[\frac{s_{1}}{2}]D_{2}[\frac{s_{2}}{2}]+B_{-}D_{1}[\frac{s_{1}}{2}]D^{\dagger}_{2}[\frac{s_{2}}{2}]\Bigg\}\vert\phi\rangle,
\end{align}
where $\kappa=1/\sqrt{P_{s}}$ is the normalization factor and $P_{s}=\langle\tilde{\Phi}\vert\tilde{\Phi}\rangle$
denotes the post-selection success probability, $A_{\pm}=(\hat{I}\pm\langle\hat{\sigma}_{x}\rangle_{w})(\hat{I}\pm\langle\hat{\sigma}_{y}\rangle_{w})$
and $B_{\pm}=(\hat{I}\mp\langle\hat{\sigma}_{x}\rangle_{w})(\hat{I}\pm\langle\hat{\sigma}_{y}\rangle_{w})$.
with the weak value (WV) $\langle\hat{\sigma}_{x}\rangle_{w}$ and
$\langle\hat{\sigma}_{y}\rangle_{w}$ given by

\begin{align}
\langle\hat{\sigma}_{x}\rangle_{w} & =\frac{\langle\psi_{f}\vert\hat{\sigma}_{x}\vert\psi_{i}\rangle}{\langle\psi_{f}\vert\psi_{i}\rangle}=e^{i\delta_{1}}\tan(\frac{\theta_{1}}{2}).\label{eq:weak-value-1}\\
\langle\hat{\sigma}_{y}\rangle_{w} & =\frac{\langle\psi_{f}\vert\hat{\sigma}_{y}\vert\psi_{i}\rangle}{\langle\psi_{f}\vert\psi_{i}\rangle}=-ie^{i\delta_{2}}\tan(\frac{\theta_{2}}{2}).\label{eq:weak-value-2}
\end{align}
Here the parameters satisfy $\delta_{j}\in[0,2\pi]$, $\theta_{j}\in[0,\pi)$
for $j=1,2$. For concreteness, we consider two separate pre-selection
settings: one for $\langle\hat{\sigma}_{x}\rangle_{w}$ and one for
$\langle\hat{\sigma}_{y}\rangle_{w}$%
. The weak values (WV) are generally complex. Equations (\ref{eq:weak-value-1})
and (\ref{eq:weak-value-2}) show that when the pre-selected $\vert\psi_{i}\rangle$
and post-selected states $\vert\psi_{f}\rangle$ are nearly orthogonal,
the weak values can become large, leading to weak-value amplification
(WVA) \citep{RevModPhys.86.307}. It is worth noting that the weak
value is inherently a product of the two-state vector formalism (TSVF)
\citep{PhysRev.134.B1410,Aharonov2008}, where the system is characterized
by both a pre-selected state evolving forward in time and a post-selected
state evolving backward. Within this framework, the weak value is
solely determined by the choice of pre- and post-selected states,
independent of the measurement strength \citep{PhysRevA.83.052106,PhysRevA.106.022619,PhysRevA.102.042601,Turek_2015}.
Hence, contrary to what the name \textquotedbl weak value\textquotedbl{}
might imply, its existence does not require a weak interaction. It
remains a well-defined quantity across the entire spectrum of measurement
strengths, from the weak-coupling limit all the way to projective
measurements \citep{PhysRevA.105.042202,PhysRevA.98.042112,Hofmann_2012}.
This understanding has been experimentally verified in various platforms,
demonstrating that weak values can be extracted even under strong
measurement conditions using appropriate projective measurements and
phase rotations \citep{Wagner_2024}.

\begin{figure}
\centering
\begin{centering}
\includegraphics[scale=0.405]{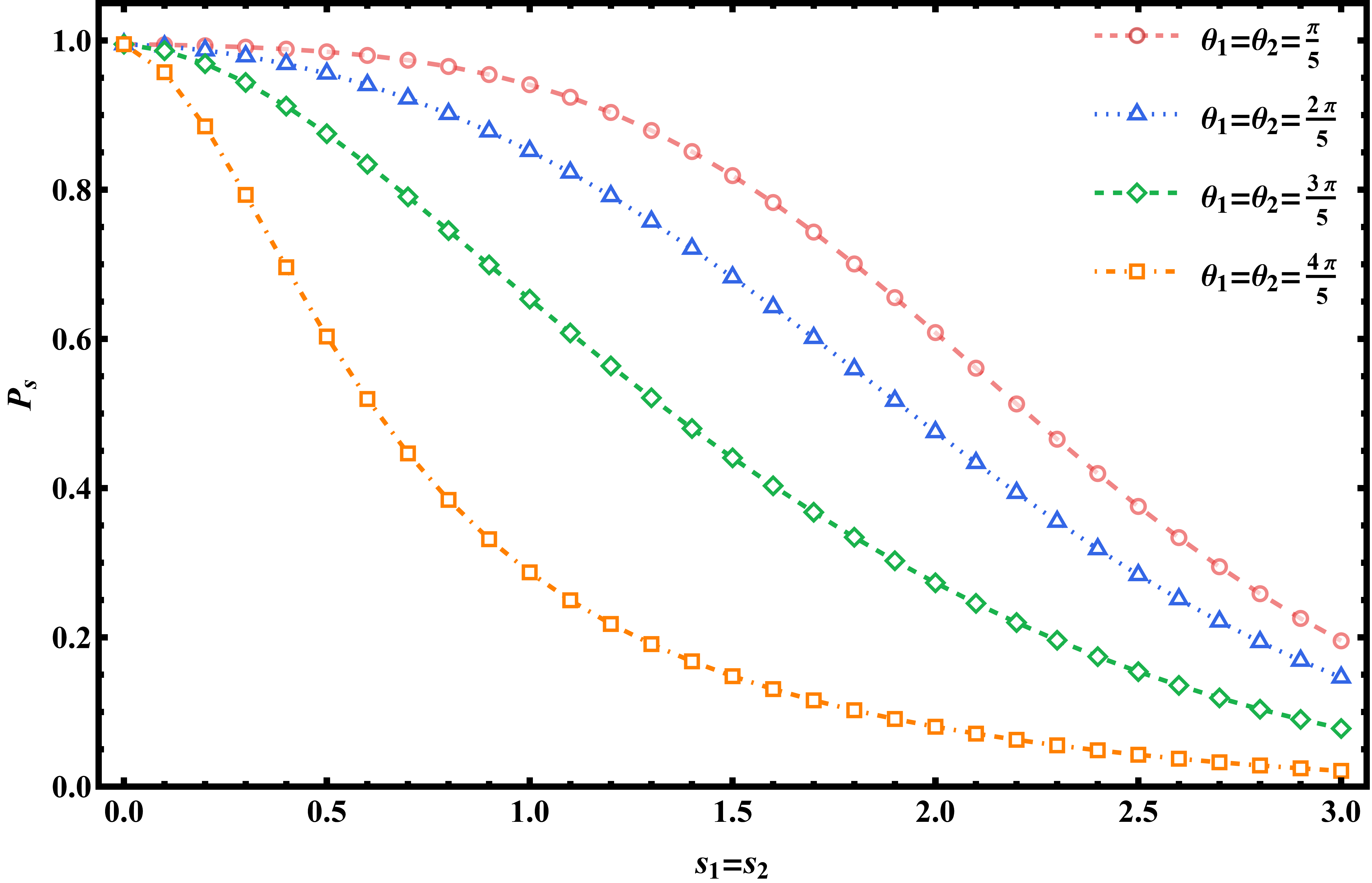} 
\par\end{centering}
\caption{Post-selection success probability $P_{s}$ of the state $\left|\Phi\right\rangle $
as a function of the coupling strength $s_{i}$ for different weak-value
parameters. Fixed parameters are $\mu=0$, $\varphi=\delta_{1}=\delta_{2}=\pi/2$,
and $r=0.1$. \label{fig:probability}}
\end{figure}

As shown in Fig. \ref{fig:probability}, the final pointer state $|\Phi\rangle$
depends on both the WV and the coupling strengths. The success probability
$P_{s}$ decreases as the weak value increases in the weak-measurement
regime. However, larger coupling coefficients $s_{1}$ and $s_{2}$
can, in some parameter regimes, partially compensate for the reduced
success probability associated with anomalous weak values, which are
typically associated with low post-selection probabilities.

It is important to emphasize that the final pointer state $|\Phi\rangle$
is obtained without using the first-order Aharonov--Albert--Vaidman
(AAV) approximation \citep{PhysRevA.102.042601,PhysRevA.83.052106}.
We retain the finite-coupling unitary evolution and evaluate the post-selected
pointer state directly. This treatment allows the parameters $s_{1}$
and $s_{2}$ to be explored in a weak-to-intermediate coupling regime,
rather than being restricted to the strict infinitesimal-coupling
limit. As a result, the weak values $\langle\hat{\sigma}_{x}\rangle_{w}$
and $\langle\hat{\sigma}_{y}\rangle_{w}$ remain the quantities determined
by the chosen pre- and post-selected states, and the post-selection
back action can be analyzed beyond the simplest AAV linear-response
picture. In summary, this work provides a general framework in which
PVNMs act as a tunable mechanism for modifying the nonclassical properties
and metrological performance of ECS.

In the following sections, we examine the specific effects that anomalous
WV of the measured system observable exert on the intrinsic properties
of $|\Phi\rangle$.

\section{EFFECTS ON SQUEEZING\label{3} }

In this section, we examine the influence of PVNMs on the squeezing
properties of ECS by considering both quadrature squeezing (QS) and
sum squeezing (SS).

\begin{figure*}
\centering
\begin{centering}
\includegraphics[scale=0.145]{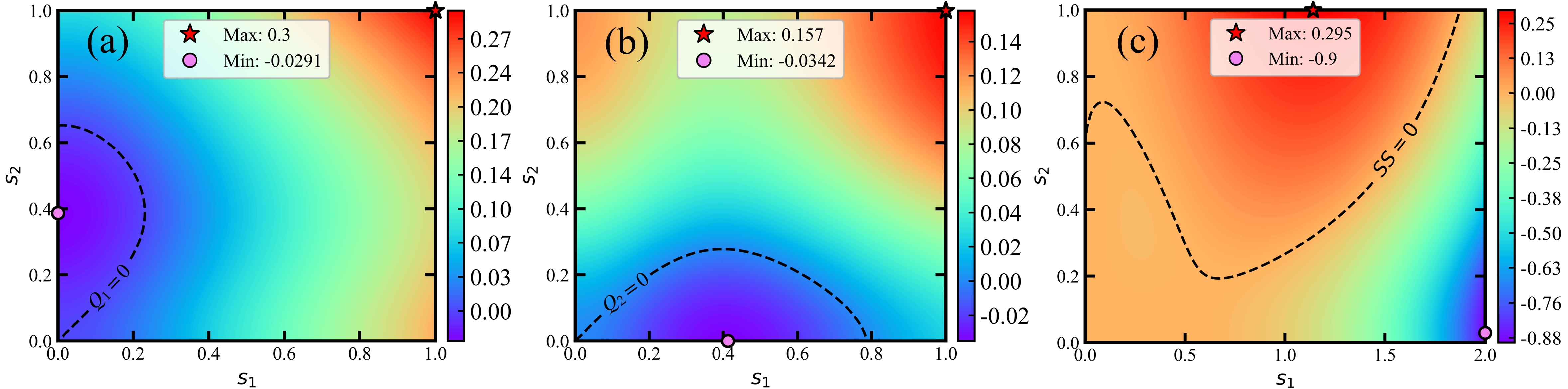} 
\par\end{centering}
\caption{Squeezing parameters in the $s_{1}-s_{2}$ plane: (a) $Q_{1}$ squeezing,
(b) $Q_{2}$ squeezing, and (c) Sum squeezing in the state $|\Phi\rangle$
as functions of the coupling strengths $s_{1}$ and $s_{2}$. Fixed
parameters are $\mu=\varphi=\Theta=\delta_{1}=\delta_{2}=\pi/2$,
$\theta_{1}=\theta_{2}=4\pi/5$, and $r=0.1$. \label{fig:squeezing}}
\end{figure*}

\subsection{Quadrature Squeezing\label{3-1} }

Squeezing is a hallmark of nonclassicality in quantum optics. It arises
from the redistribution of quadrature fluctuations, allowing one quadrature
to fall below the standard quantum limit \citep{9cxn-t1vf,app151810179}.
We investigate this effect for ECS under PVNMs.

We first introduce a pair of quadrature operators defined as 
\begin{align}
\hat{F}_{1} & =(\hat{a}+\hat{b}+\hat{a}^{\dagger}+\hat{b}^{\dagger})/2^{3/2},\\
\hat{F}_{2} & =(\hat{a}+\hat{b}-\hat{a}^{\dagger}-\hat{b}^{\dagger})/i2^{3/2},
\end{align}
where $\hat{a}$ and $\hat{b}$ represent annihilation operators for
two distinct spatial modes, with their Hermitian conjugates denoting
the corresponding creation operators. These operators characterize
a pair of orthogonal variables in phase space. They obey the commutation
relation $[\hat{F}_{1},\hat{F}_{2}]=i/2,$ which leads directly to
the uncertainty relation 
\begin{equation}
\Delta\hat{F}^{2}_{1}\Delta\hat{F}^{2}_{2}\geq1/16.
\end{equation}
Here, $\Delta\hat{F}^{2}_{i}=\langle\hat{F}^{2}_{i}\rangle-\langle\hat{F}_{i}\rangle^{2}$
denotes the quantum fluctuation of the $i$th quadrature component.
To quantify the degree of squeezing, we define a squeezing parameter
as 
\begin{equation}
Q_{i}=\Delta\hat{F}^{2}_{i}-1/4,
\end{equation}
where $1/4$ corresponds to the noise level of a coherent state, the
standard quantum limit for a minimum uncertainty state. When $-1/4<Q_{i}<0$,
the state is squeezed along that quadrature, indicating fluctuations
below the coherent-state level. The theoretical lower bound $Q_{i}=-1/4$
corresponds to perfect squeezing, where the variance vanishes entirely.
After algebraic simplification, the orthogonal squeezing parameter
$Q_{i}$ of $\vert\Phi\rangle$ can be written as

\begin{align}
Q_{1}= & \frac{1}{4}\left[\langle\hat{a}^{\dagger}\hat{a}\rangle+\langle\hat{b}^{\dagger}\hat{b}\rangle+\operatorname{Re}[\langle\hat{a}^{2}\rangle+\langle\hat{b}^{2}\rangle]\right]+\frac{1}{2}\operatorname{Re}[\langle\hat{a}\hat{b}\rangle]\nonumber \\
 & +\frac{1}{2}\operatorname{Re}[\langle\hat{a}^{\dagger}\hat{b}\rangle]-\frac{1}{2}\left[\operatorname{Re}[\langle\hat{a}\rangle]+\operatorname{Re}[\langle\hat{b}\rangle]\right]^{2}.
\end{align}
and 
\begin{align}
Q_{2}= & \frac{1}{4}\left[\langle\hat{a}^{\dagger}\hat{a}\rangle+\langle\hat{b}^{\dagger}\hat{b}\rangle-\operatorname{Re}[\langle\hat{a}^{2}\rangle+\langle\hat{b}^{2}\rangle]\right]-\frac{1}{2}\operatorname{Re}[\langle\hat{a}\hat{b}\rangle]\nonumber \\
 & -\frac{1}{2}\operatorname{Re}[\langle\hat{a}^{\dagger}\hat{b}\rangle]-\frac{1}{2}\left[\operatorname{Im}[\langle\hat{a}\rangle]+\operatorname{Im}[\langle\hat{b}\rangle]\right]^{2}.
\end{align}

Here $\langle\cdot\rangle$ denotes expectation values taken with
respect to the states $|\phi\rangle$ and $|\Phi\rangle$. A direct
calculation for $|\phi\rangle$ shows that the ECS exhibit no squeezing
in the two orthogonal directions $Q_{1,\phi}=Q_{2,\phi}=0$. We then
computed $Q_{1}$ and $Q_{2}$ for the state $|\Phi\rangle$ and plotted
their dependence on $s_{1}$ and $s_{2}$ in Fig. \ref{fig:squeezing}
to illustrate the trends. As shown in Fig. \ref{fig:squeezing}(a),
$Q_{1}<0$ in the plotted region, indicating quadrature squeezing
below the coherent-state level. More importantly, Fig. \ref{fig:squeezing}(b)
exhibits a distinct, though narrow, region of negative values at small
$s_{2}$ and intermediate $s_{1}$. That negative value corresponds
to a measurable reduction of fluctuations below the standard quantum
limit and demonstrates that phase-sensitive post-selection can produce
quadrature squeezing. The effect is directional and parameter selective;
by tuning the coupling strengths and relative phases, the squeezed
regime can be accessed reproducibly. The magnitude of the squeezing
is modest, but its rapid appearance and disappearance with $s_{1}$
and $s_{2}$ point to clear operating conditions for optimization.

Overall, the comparison between Figs. \ref{fig:squeezing}(a) and
\ref{fig:squeezing}(b) shows that post-selection induced interference
provides a practical and tunable route to generate quadrature squeezing
with modest resources, suitable for proof-of-principle demonstrations
and for applications that can exploit small, reliably produced reductions
in noise.

\begin{figure*}
\centering
\begin{centering}
\includegraphics[scale=0.045]{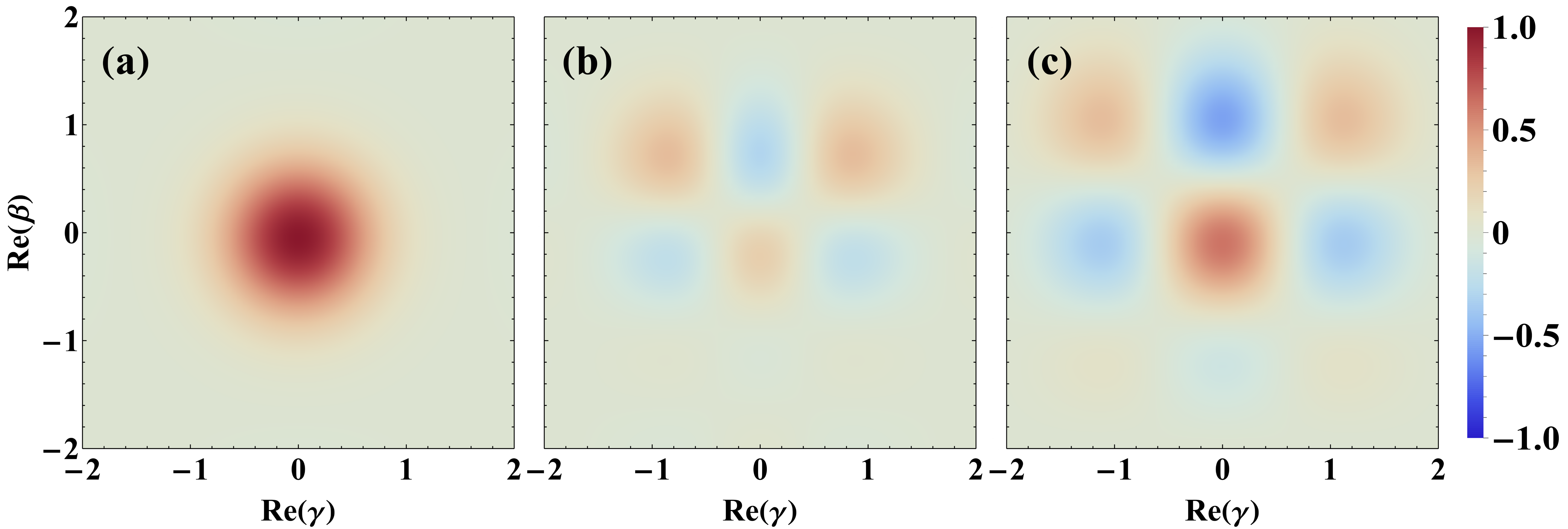} 
\par\end{centering}
\caption{(a)-(c): Real-axis cross sections ($\operatorname{Im}\gamma=\operatorname{Im}\beta=0$)
of the joint quasiprobability $P_{J}(\gamma,\beta)$ in the $\operatorname{Re}\gamma-\operatorname{Re}\beta$
plane for $s_{1}=s_{2}=0,1,2$. (d)-(f): single-mode Wigner $W(x,p)$
of the reduced state for $s_{2}=0$ and $s_{1}=0,1,2$, respectively.
Other parameters are the same as in Fig. \ref{fig:squeezing}. \label{fig:wigner}}
\end{figure*}

\subsection{Sum squeezing\label{3-2} }

In this subsection, we investigate how PVNMs influence the squeezing
properties of ECS, focusing on SS. This quantity, first introduced
by Hillery \citep{PhysRevA.40.3147}, serves as a sensitive indicator
of nonclassical correlations in two-mode radiation fields. It is defined
as follows:

\begin{equation}
V_{\Theta}=\frac{1}{2}(e^{i\Theta}\hat{a}^{\dagger}\hat{b}^{\dagger}+e^{-i\Theta}\hat{a}\hat{b}),\label{eq:quadrature operator}
\end{equation}
Here, $\Theta\in[0,2\pi]$ defines the rotation angle of $V_{\Theta}$
relative to the real axis in the complex plane \citep{NGUYENBAAN199934,PhysRevLett.112.070402,REN2019106,PhysRevA.40.3147}.
A state is said to be SS for a value of $\Theta$ if $\langle V^{2}_{\Theta}\rangle-\langle V_{\Theta}\rangle^{2}<\langle N_{a}+N_{b}+1\rangle/4$,
where $N_{a}=\hat{a}^{\dagger}\hat{a}$ and $N_{b}=\hat{b}^{\dagger}\hat{b}$,
then we can define the degree of SS in the following manner

\begin{equation}
SS=\frac{4(\langle V^{2}_{\Theta}\rangle-\langle V_{\Theta}\rangle^{2})}{\langle N_{a}+N_{b}+1\rangle}-1.\label{eq:sum squeezing}
\end{equation}

For $-1\leq SS<0$, the two-mode state exhibits squeezing, with the
squeezing strength increasing as SS approaches $-1$. Substituting
Eq. (\ref{eq:quadrature operator}) into Eq. (\ref{eq:sum squeezing})
yields an expression for SS in normally ordered operator form 
\begin{equation}
SS=\frac{2[Re[e^{-2i\Theta}\langle\hat{a}^{2}\hat{b}^{2}\rangle]-2(Re[e^{-i\Theta}\langle\hat{a}\hat{b}\rangle])^{2}+\langle N_{a}N_{b}\rangle]}{\langle N_{a}\rangle+\langle N_{b}\rangle+1}.
\end{equation}
Here, $\langle\cdot\rangle$ indicates the expectation value evaluated
in $\vert\Phi\rangle$. Setting $s_{1}=s_{2}=0$ recovers the ECS
inherent $SS=0$. As shown in Fig. \ref{fig:squeezing}(c), the total
squeezing parameter undergoes a continuous evolution from negative
to positive values, with the corresponding coupling coefficients $s_{1}$
and $s_{2}$ varying accordingly. Within the regime of weakly coupled
systems, the system primarily exhibits behaviour that is anti-squeezed.
As coupling coefficients $s_{1}$ and $s_{2}$ increase, the system
undergoes a gradual transition into an enhanced squeezing state. It
is noteworthy that when $s_{2}$ is constrained within the weaker
region, the magnitude of $s_{1}$ is found to exhibit a substantial
amplification in its quantum properties%
. This result demonstrates that quantum squeezing behaviour can be
effectively manipulated by regulating the coupling strength, enabling
flexible switching between squeezed and anti-squeezed states. This
property is of considerable significance for the preparation of nonclassical
light fields and parameter optimisation in the domain of quantum information
processing. To further support these conclusions, the ensuing discourse
employs the joint Wigner function as a fundamental tool for validation.

\section{Measurement Backaction in Phase Space\label{4} }

\subsection{Joint Wigner function\label{4-1}}

To investigate the effect of PVNMs on the initial pointer state $\vert\phi\rangle$,
we analyze the phase-space distribution of the post-measurement state
$\vert\Phi\rangle$ through its Wigner function. The joint Wigner
function of the two-mode state $\vert\Phi\rangle$ is defined as

\begin{align}
W_{J}\left(\gamma,\beta\right) & =\frac{4}{\pi^{2}}\operatorname{Tr}[\rho_{ab}D(\gamma)D(\beta)P_{j}D^{\dagger}(\beta)D^{\dagger}(\gamma)]\nonumber \\
 & =\frac{4}{\pi^{2}}P_{J}(\gamma,\beta),\label{eq:joint wigner functuon}
\end{align}
where $\rho_{ab}=\vert\Phi\rangle\langle\Phi\vert$ is the density
operator of state $\vert\Phi\rangle$, and%
{} $P_{j}=P_{a}P_{b}=\exp[i\pi a^{\dagger}a+i\pi b^{\dagger}b]$ is
the joint photon number parity operator. The operators $D(\gamma)=\exp[\gamma a^{\dagger}-\gamma^{*}a]$
and $D(\beta)=\exp[\beta b^{\dagger}-\beta^{*}b]$ are the displacement
operators acting on the $a$ and $b$-modes of the two-mode state
$\vert\Phi\rangle$, respectively. The joint Wigner function $W_{J}$
describes the state in a four-dimensional (4D) phase space with coordinates
($\text{Re}(\gamma)$, $\text{Im}(\gamma)$, $\text{Re}(\beta)$,
$\text{Im}(\beta)$). For the phase-space analysis of the state $\vert\Phi\rangle$,
we use the scaled Wigner function $P_{J}(\gamma,\beta)$, with

\begin{equation}
-1\leq P_{J}(\gamma,\beta)\leq1.
\end{equation}

To highlight the key characteristics of the 4D Wigner function for
the state $|\Phi\rangle$ and to compare it with the initial state
$|\phi\rangle$, we begin by presenting 2D cross-sections along the
$\text{Re}(\gamma)$ and $\text{Re}(\beta)$ plane of the joint Wigner
function. From Figs. \ref{fig:wigner}(a), \ref{fig:wigner}(b), and
\ref{fig:wigner}(c), it can be observed that the joint Wigner function
$P_{J}(\gamma,\beta)$ exhibits significant structural evolution with
variations in the coupling strength. When $s_{1}=s_{2}=0$, the distribution
assumes a concentrated symmetric form with a limited range of negative
values, indicating weaker nonclassical properties in the system. As
the parameters increase to $s_{1}=s_{2}=1$, the joint distribution
develops a multi-branch structure with a markedly enhanced negative
region, indicating a pronounced quantum interference effect. Under
strong coupling conditions $s_{1}=s_{2}=2$, the negative region further
expands to form intricate interference fringes, reflecting the system's
highly nonclassical behaviour and multi-mode quantum correlations.

Given that the joint Wigner function faithfully reflects the underlying
structure%
{} of a quantum state, it is well suited for probing changes induced
by measurement. From this perspective, a detailed examination of how
the phase-space distribution of the a %
mode evolves in the pointer state $\vert\Phi\rangle$ after PVNMs
becomes essential for clarifying how the intrinsic features of the
ECS are affected.

\subsection{Wigner function\label{4-2}}

The Wigner function for the reduced single-mode state $\hat{\rho}_{a}=\operatorname{Tr}_{b}[|\Phi\rangle\langle\Phi|]$
is given by \citep{Int}

\begin{equation}
W(x,p)=\frac{1}{\pi^{2}}\iint^{\infty}_{-\infty}e^{2i(p\lambda^{\prime}-x\lambda^{\prime\prime})}\operatorname{Tr}[\rho_{a}e^{\lambda\hat{a}^{\dagger}-\lambda^{*}\hat{a}}]d\lambda^{\prime}d\lambda^{\prime\prime},\label{eq:wigner-function}
\end{equation}
where, $\lambda=\lambda'+i\lambda''$ , $x$, and $p$ to emphasize
the analogy between the quadratic radiation field and the normalized
dimensionless position and momentum observables of the beam in phase
space. 

Figures \ref{fig:wigner}(d), \ref{fig:wigner}(e), and \ref{fig:wigner}(f)
show the Wigner function $W(x,p)$ for different coupling strengths
$s_{1}$. As $s_{1}$ increases, the Wigner function peak shifts from
phase space center to the edge, with shape becoming increasingly irregular.
Compared with the initial state, the post-measurement state develops
negative regions in phase space and exhibits a distinct squeezing
effect. Both this squeezing effect and the nonclassicality (evidenced
by larger and more pronounced negative regions) strengthen as $s_{1}$
increases. This indicates that, under appropriate parameters, the
measurement significantly enhances the nonclassical characteristics
of the pointer state.

Overall, this visualization shows that modulating the coupling strength
significantly modifies the nonclassical phase-space features of the
pointer state. Building on these results, we employ the Hillery--Zubairy
criterion and linear entropy to characterize the influence of PVNMs
on quantum correlations in ECS.

\section{Effect of entanglement\label{5}}

In this section, we study how PVNMs modify the quantum correlations
of ECS. In particular, we analyze the post-measurement Hillery--Zubairy
correlations and the linear-entropy coefficients that appear in several
multimode entanglement criteria to elucidate how the measurement process
alters quantum-correlation properties.

\begin{figure*}[t]
\centering
\begin{centering}
\includegraphics[scale=0.08]{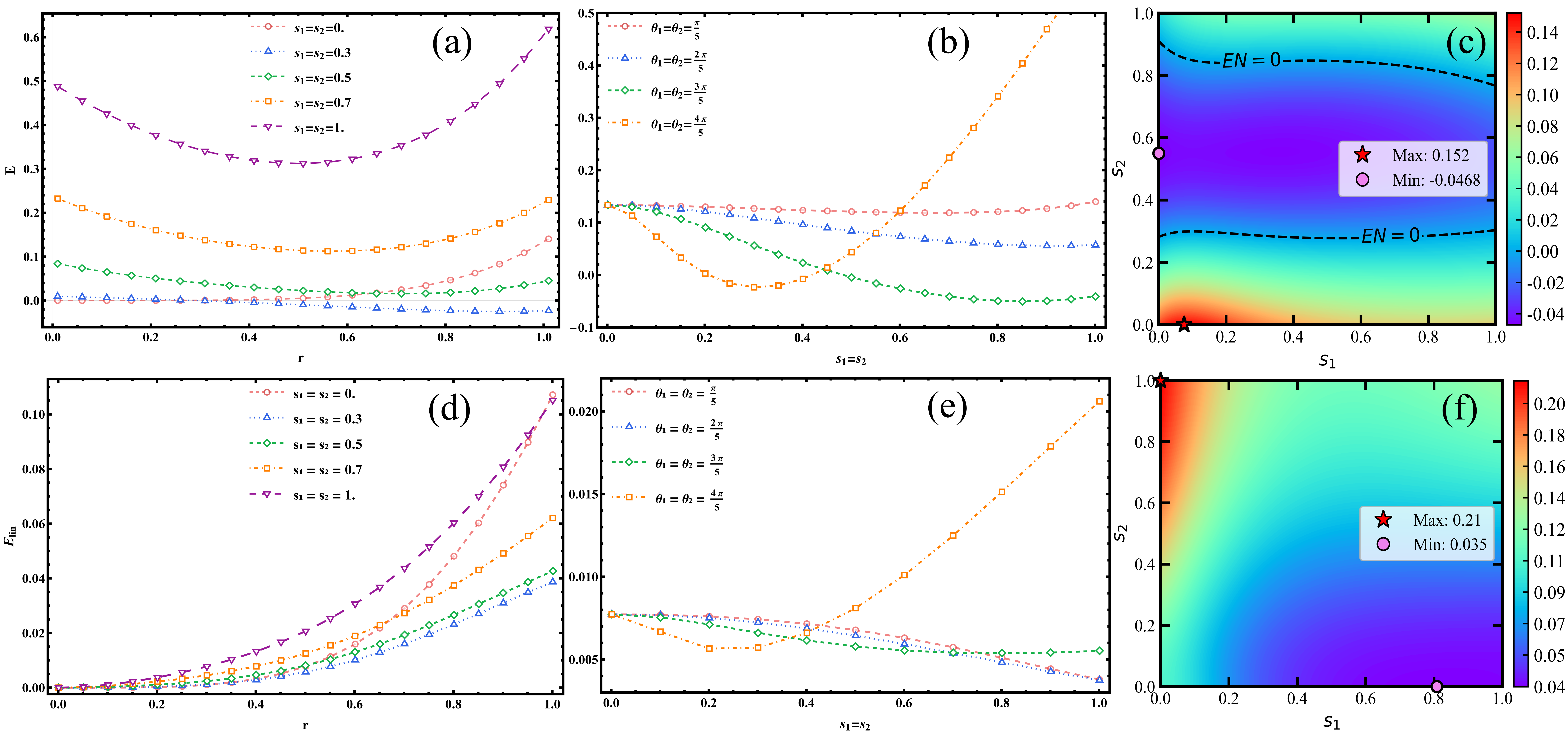} 
\par\end{centering}
\caption{Hillery-Zubairy correlation $E$ and linear entropy $E_{\mathrm{lin}}$for
the state $|\Phi\rangle$. Panels (a) and (d) show $E$ and $E_{\mathrm{lin}}$
as functions of $r$ for different symmetric couplings $s_{1}=s_{2}$.
Panels (b) and (e) show $E$ and $E_{\mathrm{lin}}$ as a function
of $s_{1}=s_{2}$ for several weak-value parameters at $r=0.5$. Panels
(c) and (f) show $E$ and $E_{\mathrm{lin}}$ on the $s_{1}$ and
$s_{2}$ plane at $r=1$. Other parameters are the same as in Fig.
\ref{fig:squeezing}. \label{fig:entanglement}}
\end{figure*}

\subsection{Hillery-Zubairy correlation\label{5-1}}

To characterize the entanglement between the two modes of the optical
field, we begin by employing the Hillery-Zubairy correlation (HZC)
\citep{PhysRevLett.96.050503}, defined as follows \citep{dao2015quantum}:

\begin{align}
E & =\langle\hat{a}^{\dagger}\hat{a}\rangle\langle\hat{b}^{\dagger}\hat{b}\rangle-|\langle\hat{a}\hat{b}\rangle|^{2}.\label{eq:entanglement-Eq}
\end{align}
This entanglement criterion underscores the significance of the correlation
$\langle\hat{a}\hat{b}\rangle$ between the two modes of a state.
Negative values of $E$ certify entanglement, whereas positive values
do not imply separability. Additionally, the relation $\vert\langle\hat{a}\hat{b}\rangle\vert^{2}\le\langle\hat{a}^{\dagger}\hat{a}\rangle\langle\hat{b}\hat{b}^{\dagger}\rangle$
\citep{PhysRevResearch.3.033095,PhysRevLett.96.050503} holds, yielding
the bound that the entanglement condition for two-mode fields is restricted
to $-\langle a^{\dagger}a\rangle\le E<0$. For the initial ECS $\vert\phi\rangle$
the HZC gives $E_{\phi}=0.25\vert\alpha\vert^{2}(1+\exp[-\vert\alpha\vert^{2}])^{-2}$,
the HZC quantity depends on the coherent amplitude $|\alpha|$. Since
the HZC serves as an entanglement witness, its value should be interpreted
together with its sign%
: negative values certify entanglement, whereas positive values alone
do not imply separability.

We analyze compliance with the HZC for the final pointer state $|\Phi\rangle$
by computing Eq. (\ref{eq:entanglement-Eq}). As shown in Fig.\ref{fig:entanglement},
the HZC reveals that the post-selected two-mode entanglement depends
sensitively on the WV parameters. The two-dimensional density plot
in Fig. \ref{fig:entanglement}(a) shows that when the coupling parameters
$s_{1}$ and $s_{2}$ increase synchronously, the criterion quantity
$E$ systematically deviates from the classical limit along the $s_{1}=s_{2}$
direction, indicating that nonclassical correlations are enhanced
over a wide parameter domain. Fig. \ref{fig:entanglement}(b) presents
the quantitative behavior of $E$ as the auxiliary parameter s increases
under several fixed coupling strengths $s_{1}=s_{2}$, with larger
$s_{1,2}$ corresponding to steeper response curves, indicating that
WVA can increase the amplitude of the entanglement witness and enhance
its sensitivity to changes in system parameters. Fig. \ref{fig:entanglement}(c)
shows that the WV parameters $\theta_{1}$ and $\theta_{2}$ can effectively
modulate the amplification amplitude. Specific phase combinations
can make the deviation of the criterion more significant, while other
phases suppress this effect. In summary, weak measurement and post-selection
can redistribute interference weights and thereby modify the observed
bipartite quantum correlations. This can enhance entanglement signatures
while maintaining continuous control within experimentally accessible
coupling and phase parameters \citep{PhysRevResearch.2.023047}. The
approach enhances entanglement detectability and offers a controllable
route for ECS-based quantum information processing and quantum metrology.
These results are further corroborated by linear entropy analysis.

\subsection{Linear entropy \label{5-2} }

Linear entropy serves as a convenient measure of bipartite entanglement
\citep{PhysRevLett.96.050503}, grounded in the established relation
between entanglement and the mixedness of reduced states \citep{Agarwal_2005,PhysRevA.80.020101,Chuong_2023}.
Formally, the linear entropy of a subsystem is given by:

\begin{equation}
E_{\mathrm{lin}}=1-\operatorname{Tr}\bigl(\rho^{2}_{a}\bigr)=1-\operatorname{Tr}\bigl(\rho^{2}_{b}\bigr),
\end{equation}
where $\rho_{a}=\operatorname{Tr}_{b}[|\Phi\rangle\langle\Phi|]$
and $\rho_{b}=\operatorname{Tr}_{a}[|\Phi\rangle\langle\Phi|]$ denote
the reduced density matrices obtained by partial trace over mode~$a$
and mode~$b$. This quantity is bounded between $0$ and $1$, $E_{\mathrm{lin}}=0$
corresponds to a pure reduced state, indicating separability, while
$E_{\mathrm{lin}}\to1$ signifies a maximally mixed reduced state,
indicating stronger entanglement within the explored parameter regime.
Unlike the von Neumann entropy, the linear entropy circumvents the
need for diagonalizing the density matrix and can be interpreted as
its leading (quadratic) approximation, making it a particularly practical
tool for the analysis of non-Gaussian states.

As shown in Figs. \ref{fig:entanglement}(d), \ref{fig:entanglement}(e),
and \ref{fig:entanglement}(f),%
{} the linear entropy $E_{\mathrm{lin}}$ employed as a measure of two-mode
entanglement is plotted quantitatively against the system parameter
$r$ for several fixed values of the symmetric coupling strength parameter
$s_{1}=s_{2}$. The analysis reveals that, for any given coupling
strength parameter $s_{1}$ and $s_{2}$, the linear entropy $E_{\mathrm{lin}}$
increases monotonically with $r$, confirming that $r$ is the dominant
physical parameter driving the generation and growth of quantum entanglement
in this setting.

More importantly, at fixed value of $r$, the linear entropy $E_{\mathrm{lin}}$
increases as the coupling strengths $s_{1}$ and $s_{2}$ are increased.
In particular, when $s_{1}=s_{2}=1$,%
{} the system exhibits the highest degree of entanglement across the
whole parameter interval $0\leq r<1$. This observation suggests that,
in addition to the entanglement-generation mechanism governed by $r$,
applying symmetric and tunable coupling strengths $s_{1}$ and $s_{2}$
to both modes can enhance the final entanglement resource. Our results
provide both a theoretical basis and an experimentally feasible protocol
for on-demand control of entanglement in continuous-variable systems
\citep{PhysRevA.108.012432}, realized through composite parameter
tuning and squeezing operations.

\begin{figure*}
\centering
\begin{centering}
\includegraphics[scale=0.175]{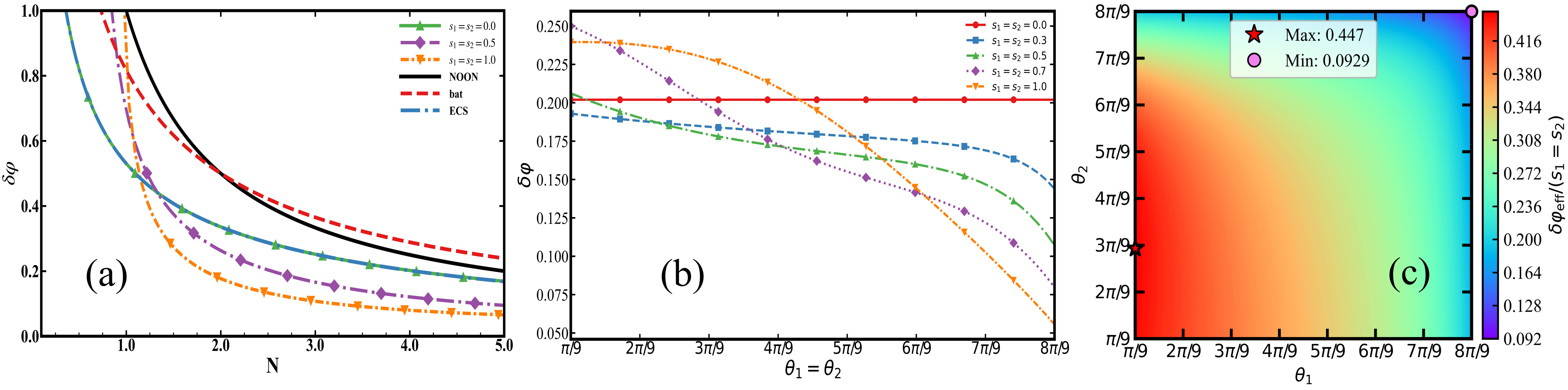} 
\par\end{centering}
\caption{Quantum Cramér-{}-Rao bound $\delta\varphi$ for the state $|\Phi\rangle$.
(a) $\delta\varphi$ versus $N$ for $s_{1}=s_{2}=0,0.5,1$, and $\theta_{1}=8\pi/9,\ \theta_{2}=\pi/9$,
with NOON, bat, and ECS states as references. (b) $\delta\varphi$
versus $\theta_{1}=\theta_{2}$ for $s_{1}=s_{2}=0$ to $1$. (c)
$\delta\varphi_{\mathrm{eff}}$ versus $\theta_{1},\theta_{2}$ at
$s_{1}=s_{2}=0.5$. Fixed parameters are $\mu=\varphi=\delta_{1}=\delta_{2}=\pi/2$.
\label{fig:QCRB}}
\end{figure*}

\section{Cramér--Rao bound\label{6} }

The metrological relevance of post-selected weak measurement lies
in whether the post-selected state can improve phase-estimation precision.
To investigate this, we employ the Quantum Fisher Information (QFI)
as our core tool. Entangled coherent states have been shown to provide
superior phase sensitivity over NOON and bat states under both ideal
and lossy conditions \citep{PhysRevLett.107.083601}. In that standard
ECS metrology, the original ECS outperforms NOON and bat states in
the small-photon-number regime and approaches the NOON limit for large
photon numbers. Here, we examine whether PVNMs can further enhance
this performance. As will be shown, the answer is affirmative in the
large-average-photon-number regime, albeit with a distinct parameter
dependence. The QFI quantifies the ultimate precision limit for estimating
an unknown parameter encoded in a quantum state. In the case of a
pure quantum state, it is defined as

\begin{equation}
\mathcal{F}=4[\langle\Phi'\vert\Phi'\rangle-\vert\langle\Phi'\vert\Phi\rangle\vert^{2}],
\end{equation}
with 
\begin{equation}
\vert\Phi'\rangle=\partial\vert\Phi\rangle/\partial\varphi=\mathcal{N}\chi\alpha e^{i\varphi}\hat{b}^{\dagger}|0\rangle_{a}|\alpha e^{i\varphi}\rangle_{b},
\end{equation}
where $\chi=\kappa\{t_{+}D_{1}[s_{1}/2]D_{2}[s_{2}/2]+t_{-}D^{\dagger}[s_{1}/2]D^{\dagger}[s_{2}/2]+t'_{+}D^{\dagger}[s_{1}/2]D[s_{2}/2]+t'_{-}D[s_{1}/2]D^{\dagger}[s_{2}/2]\}/4$.
In addition to ECS, we examine the performance of NOON state ($|\psi_{N}\rangle$)
and bat state ($|\psi_{B}\rangle$) \citep{eq}. The uncertainty in
estimating the parameter $\varphi$ is lower-bounded by the Quantum
Cramér-Rao bound (QCRB):

\begin{equation}
\delta\varphi\geq\frac{1}{\sqrt{\mu\mathcal{F}}}.
\end{equation}
Here, $\mathbb{\mu}$ represents the total number of independent measurements,
where $\mathbb{\mu}=1$ corresponds to a single-shot experiment \citep{PhysRevLett.105.180402,PhysRevA.81.043624}.
For a fair comparison of phase-estimation performance, we take the
total mean photon number $N$ as the resource metric. For a two-mode
state%
, the mean photon number in each mode is 

\begin{equation}
\langle\psi_{k}|a^{\dagger}a|\psi_{k}\rangle=N/2=\mathcal{N}^{2}|\alpha|^{2}.
\end{equation}
 Here, we consider four representative states, denoted by $|\psi_{k}\rangle$
with $k=N,B,E$ and $|\Phi\rangle$, such that the total mean photon
number of the system is $N$, the phase uncertainty $\delta\varphi$
becomes a function of $N$. Considering the situation with no loss,
the optimal phase estimation of the pure states is analytically soluble,
yielding $\delta\varphi_{N}\geq1/N$ for the NOON state, $\delta\varphi_{B}\geq1/\sqrt{N(N/2+1)}$
for the bat state, and for the ECS $\delta\varphi_{E}\geq1/2\alpha\mathcal{N}\{1+[1-(\mathcal{N}\alpha)^{2}]\alpha^{2}\}^{1/2}$,
thereby enabling a direct resource-based comparison, with detailed
derivations given in Appendix \ref{sec:A1}.

To quantify the overall metrological performance beyond the conditional
post-selected channel, it is useful to introduce an effective figure
of merit that accounts for the post-selection probability. The QFI
and QCRB discussed above refer to the successful post-selected subensemble.
Therefore, the apparent metrological enhancement should be interpreted
together with the post-selection success probability $P_{s}$, since
only a fraction of the initial trials contributes to the accepted
outcomes. A natural effective figure of merit for the full protocol
is the success-probability-weighted QFI, $\mathcal{F}_{\mathrm{eff}}=P_{s}\mathcal{F}_{\mathrm{post}}$,
which leads to the effective bound

\begin{equation}
\delta\varphi_{\mathrm{eff}}\ge\frac{1}{\sqrt{\mu\mathcal{F}_{\mathrm{eff}}}}.
\end{equation}

This effective figure of merit clarifies the trade-off between metrological
enhancement and probabilistic post-selection: a larger conditional
QFI does not necessarily imply a better overall protocol unless the
success probability remains sufficiently high. A complete resource
accounting therefore requires the joint optimization of $F_{\mathrm{post}}$
and $P_{s}$.

As shown in Fig. \ref{fig:QCRB}(a), it is instructive to compare
the performance of our post-selected ECS with that of the standard
ECS studied in \citep{PhysRevLett.107.083601}. In that work, the
original ECS was shown to outperform NOON and bat states in the small\nobreakdash-photon\nobreakdash-number
regime and to approach the NOON limit for large photon numbers. Our
scheme exhibits a distinctly different trend. As can be seen from
Fig.\,\ref{fig:QCRB}(a), for very small average photon numbers the
post-selected state does not offer a clear advantage over NOON or
bat states. However, for large average photon numbers, the post-selected
ECS achieves a significantly lower QCRB than all reference states,
including the original ECS, NOON, and bat states. This crossover indicates
that the WV\nobreakdash-induced reshaping of the pointer state becomes
effective only above a threshold photon number, where the post-selection
backaction and the associated increase in QFI overcome the intrinsic
limitations of the bare ECS. Therefore, PVNMs provide a tunable enhancement
mechanism that is most beneficial in the large\nobreakdash-average\nobreakdash-photon\nobreakdash-number
regime, complementing the advantages of standard ECS metrology. In
Fig. \ref{fig:QCRB}(b) shows that $\delta\varphi$ decreases as $\theta_{1}=\theta_{2}$
increases, and the improvement becomes more pronounced for larger
equal coupling strengths. When $s_{1}=s_{2}=0$, the bound remains
almost unchanged, whereas nonzero coupling introduces a clear dependence
on $\theta_{1}=\theta_{2}$, with the strongest coupling yielding
the best phase sensitivity over most of the range. Fig. \ref{fig:QCRB}(c)
illustrates the effective QCRB in the $\theta_{1}$ and $\theta_{2}$
plane for $s_{1}=s_{2}=0.5$. The phase sensitivity varies significantly
with both angles, with larger values of $\theta_{1}$ and $\theta_{2}$
leading to a smaller $\delta\varphi_{\mathrm{eff}}$. The maximum
and minimum values are marked by the star and circle, respectively,
indicating the parameter region that optimizes the successful post-selected
precision.

In summary, we establish the ultimate precision of a phase-estimation
protocol using PVNMs by applying the QFI and the QCRB. Our analysis
reveals that optimizing the ECS can produce a quantum gain similar
to that associated with WVA, thereby improving phase sensitivity within
the studied regime. This approach reduces the estimation error in
the successful post-selected subensemble, suggesting improved phase
sensitivity within the parameter regime considered.

\section{Fidelity\label{7}}

\begin{figure}
\centering
\begin{centering}
\includegraphics[scale=0.4]{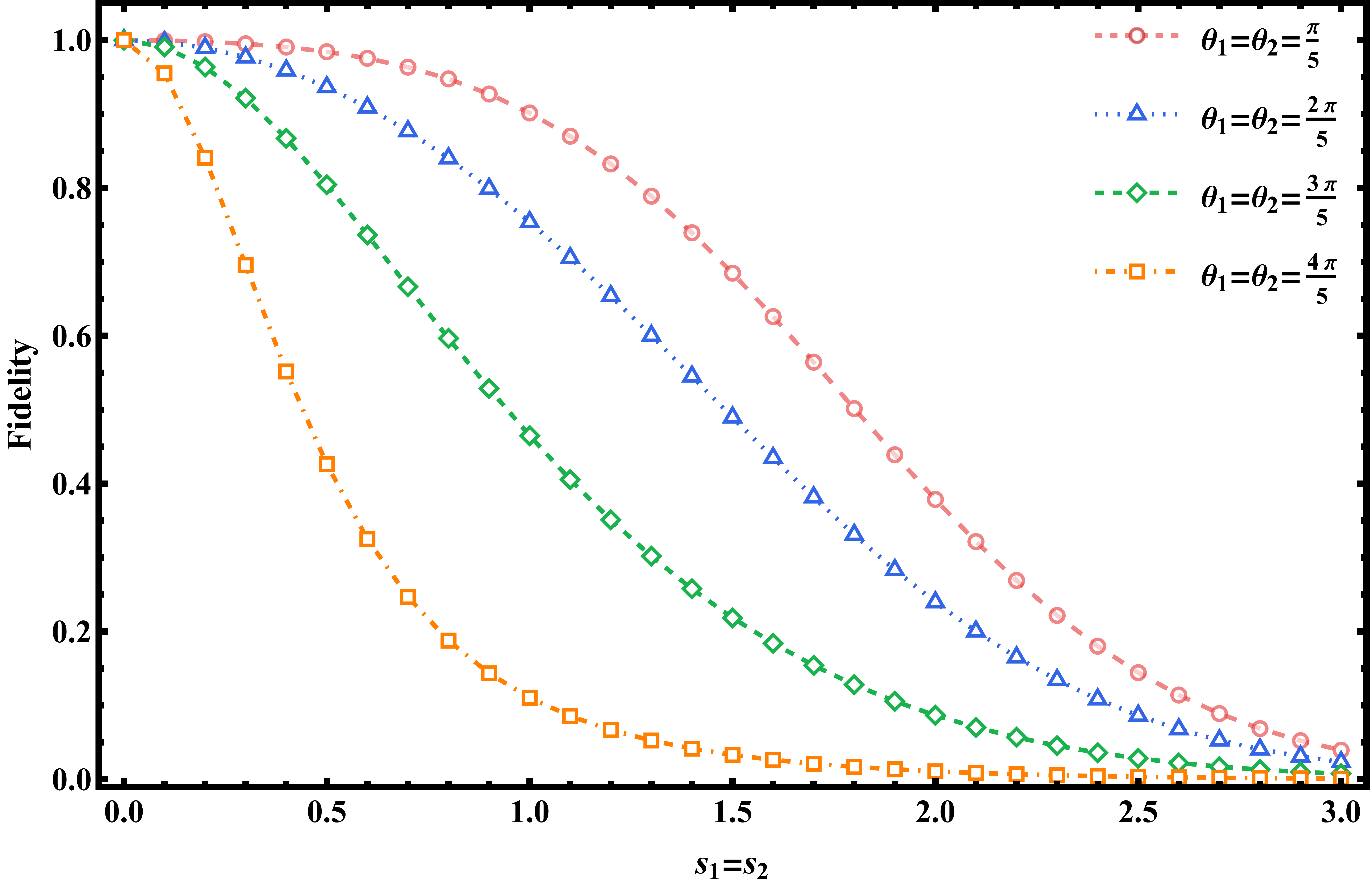} 
\par\end{centering}
\caption{Fidelity $F=|\langle\phi|\Phi\rangle|^{2}$ as a function of $r$
for different values of $s_{1}=s_{2}$. Other parameters are the same
as in Fig. \ref{fig:squeezing}.\label{fig:Fidelity}}
\end{figure}

As discussed above, PVNMs modify the quantum state and thereby influence
the nonclassical properties of ECS. To quantify the measurement-induced
disturbance, we evaluate the fidelity between the initial pointer
state $|\phi\rangle$ and the final state $|\Phi\rangle$. Expressed
as: 
\begin{equation}
F=|\langle\phi|\Phi\rangle|^{2},
\end{equation}
which satisfies $0\leq F\leq1$, with $F=1$ ($F=0$) corresponding
to identical (orthogonal) states. As shown in Fig. \ref{fig:Fidelity},
the fidelity decreases monotonically as the dimensionless coupling
strengths $s_{1}$ and $s_{2}$ increase. This behavior reflects an
increasing measurement backaction induced by stronger system--pointer
coupling, leading to a progressive departure of the post-selected
state from the initial pointer configuration. Notably, the behavior
holds generally across different WV. However, when the pre- and post-selected
states approach orthogonality, anomalously large WV arise.

In particular, when the pre- and post-selected states become nearly
orthogonal, giving rise to anomalously large WV%
, the fidelity decays significantly faster. This behavior highlights
the intrinsic trade-off of WVA: enhanced signal sensitivity is accompanied
by increased measurement-induced disturbance. This enhanced distinguishability
underscores the role of anomalous WV in amplifying the measurement-induced
disturbance, leading to a more pronounced departure of the pointer
state from its initial configuration. The observed acceleration further
illustrates the sensitivity of the measurement back-action to the
choice of post-selection, highlighting a key controllability aspect
within weak measurement frameworks. Next, we investigate the effects
of von Neumann measurement on the quantum state evolution of ECS.

\section{Time Evolution and Negativity\label{8}}

To frame this analysis, we begin by reviewing the general description
of quantum dynamics for a composite system. Typically, such a system
is initially prepared in a separable state of the form 
\begin{equation}
\rho_{in}=|\psi_{i}\rangle\langle\psi_{i}|\otimes|\phi\rangle\langle\phi|,
\end{equation}
where the subsystem and probe are uncorrelated at $t=0$. Its unitary
evolution follows \citep{quantum7020023} 
\begin{equation}
\rho_{\mathrm{after}}=U(t)\rho_{in}U^{\dagger}(t),
\end{equation}
where $U(t)$ is the evolution operator defined in Sec. \ref{2}.
After the system-pointer interaction, applying the post-selection
operator $\Pi=|\psi_{f}\rangle\langle\psi_{f}|\otimes\mathrm{I}$
yields the unnormalized pointer density operator $\tilde{\rho}=\Pi\rho_{\mathrm{after}}\Pi$,
whose trace gives the post-selection success probability $\operatorname{Tr}[\tilde{\rho}]$,
and the normalized post-selected state is 
\begin{equation}
\rho(\tau=0)=\frac{\tilde{\rho}}{\operatorname{Tr}[\tilde{\rho}]}.
\end{equation}
Here, $\tau$ represents the physical evolution time, and $\rho(\tau=0)$
defined the initial condition. To systematically study the time evolution
of the density matrix, we introduce the Lindblad master equation under
the interaction picture\citep{10.1063/1.5115323,sander2025large,vnhp-fnl3},
expressed as: 
\begin{equation}
\partial_{\tau}\rho(\tau)=\mathcal{L}[\rho(\tau)]=\mathcal{L}_{H_{\mathrm{int}}}[\rho(\tau)]+\mathcal{L}_{D}[\rho(\tau)].
\end{equation}
Here, the Liouvillian superoperator%
{} $\mathcal{L}$ is composed of the interaction Hamiltonian $\mathcal{L}_{H_{\mathrm{int}}}[\rho(\tau)]$
and dissipation $\mathcal{L}_{D}[\rho(\tau)]$. Hamiltonian evolution
term: 
\begin{equation}
\mathcal{L}_{H_{\mathrm{int}}}[\rho(\tau)]=-i[H_{\mathrm{int}},\rho(\tau)],
\end{equation}
and dissipation term: 
\begin{equation}
\mathcal{L}_{D}[\rho(\tau)]=\sum_{m=a,b}\gamma_{m}(L_{m}\rho(\tau)L^{\dagger}_{m}-\frac{1}{2}\{L^{\dagger}_{m}L_{m},\rho(\tau)\}),\label{dissipation}
\end{equation}
where $\{\hat{A},\hat{B}\}=\hat{A}\hat{B}+\hat{B}\hat{A}$ denotes
the anticommutator. We consider two optical modes independently coupled
to Markovian loss channels, with jump operators $L_{a}=\sqrt{\gamma_{a}}\hat{a}$
and $L_{b}=\sqrt{\gamma_{b}}\hat{b}$ , where $\gamma_{a}$ and $\gamma_{b}$
denote the dissipation rates of modes a and b, respectively. These
operators describe photon loss processes in the two modes and model
the coupling between the ECS and its surrounding environment, with
detailed derivations given in Appendix \ref{sec:A2}.

Within this dynamical framework, we investigate how environmental
dissipation influences the evolution of the ECS and its entanglement
properties. In particular, the resulting density matrix is used to
evaluate the entanglement negativity, allowing us to characterize
the robustness of the ECS under realistic loss conditions.

Negativity is a widely used entanglement measure for bipartite systems,
defined in terms of the partial transpose of the density matrix. For
a bipartite system $\hat{\rho}_{ab}$, the negativity is defined as
\begin{equation}
N_{e}=\frac{\|\rho^{T_{a}}_{ab}\|_{1}-1}{2},
\end{equation}
Here, $\rho^{T_{a}}_{ab}$ represents the partial transpose of mode
$a$, and $\|\cdot\|_{1}$ denotes the trace norm. A larger value
indicates stronger entanglement, while zero indicates no detectable
entanglement.

\begin{figure}
\centering
\begin{centering}
\includegraphics[scale=0.33]{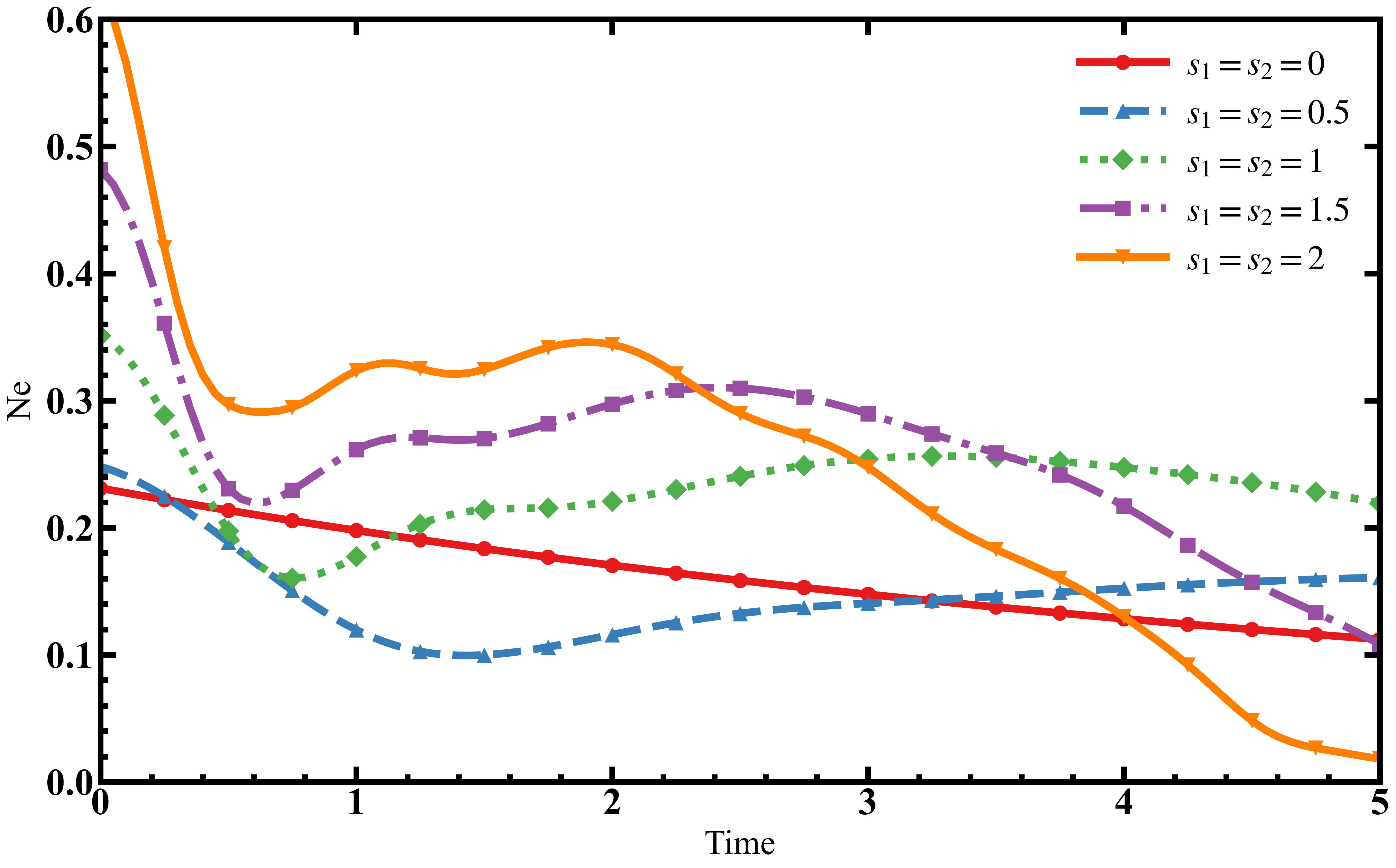} 
\par\end{centering}
\caption{Time evolution of the negativity $N_{e}$ for different coupling strengths
$s_{1}=s_{2}$. Here, $\gamma_{a}=\gamma_{b}=0.1$. Other parameters
are the same as in Fig. \ref{fig:squeezing}.\label{fig:negativity}}
\end{figure}

As shown in Fig.\ref{fig:negativity}, the dynamics reveal an interplay
between the initial entanglement and the coupling-induced modification
of entanglement. When $s_{1}=s_{2}=0$, owing to the dynamical evolution
of the open quantum system, the negativity shows a slight reduction
before stabilizing near $0.18\pm0.3$,%
. In the absence of coupling, the subsystems evolve independently,
and the initial entanglement is preserved, producing a horizontal
line. For $s_{1}=s_{2}=0.5$, the entanglement starts at 0.25 and
exhibits slow, small-amplitude damped oscillations. The weak coupling
is insufficient to generate substantial new entanglement and instead
perturbs the initial state, leading to slight reductions and fluctuations.
At $s_{1}=s_{2}=1$, the dynamics develop large-amplitude oscillations,
showing that the coupling has become strong enough to drive pronounced
entanglement oscillations, further increases in $s_{1}$ and $s_{2}$
shift the oscillation mean upward. In particular, for $s_{1}=s_{2}=2$,
the entanglement remains at a high level for most of the evolution,
indicating that strong coupling both produces additional entanglement
and strongly perturbs the initial entanglement, yielding rapid, high-intensity
dynamical behavior.

Overall, building upon the existing entanglement in the system of
ECS, increasing the coupling strength transforms the system from a
static state to large-amplitude, sustained oscillations, with both
the oscillation frequency and the average entanglement level increasing
with stronger coupling%
.

\section{CONCLUSION AND REMARKS\label{9} }

In this work, we investigate %
how post-selected measurement influences the nonclassical properties
of an entangled coherent state. Focusing on post-selected von Neumann
measurements applied to two-mode entangled coherent states, we develop
a theoretical framework to analyze how measurement back-action affects
their quantum properties. By tuning the system--pointer coupling
strength, we show that the measurement process can modify nonclassical
features and, in selected parameter regimes, enhance squeezing, Wigner
negativity, and bipartite correlations. These changes are accompanied
by improved phase-estimation performance in the successful post-selected
channel, as indicated by increased quantum Fisher information and
a reduced quantum Cramér--Rao bound, while also revealing a trade-off
with state fidelity. Notably, unlike the standard ECS metrology which
excels in the small-photon-number regime \citep{PhysRevLett.107.083601},
our post-selected scheme delivers its phase-sensitivity advantage
specifically for large average photon numbers. Our results suggest
that measurement back-action can serve as a useful mechanism for engineering
quantum resources in continuous-variable systems and provide a framework
for exploring quantum sensing protocols beyond conventional measurement
schemes.

Future work should further examine the trade-off among post-selection
success probability, state fidelity, and quantum Fisher information
in order to identify parameter regimes in which a clear metrological
advantage may emerge \citep{PhysRevLett.112.040406,PhysRevLett.115.120401}.
In parallel, it is important to extend the present framework to non-Gaussian
and hybrid quantum states, where measurement-induced effects may further
enhance nonclassical features and sensing performance \citep{62ks-19fs}.
In addition, integrating PVNMs with quantum control and error mitigation
strategies, such as coherent feedback and reservoir engineering, may
provide a route to improve the stability and robustness of the enhanced
quantum states in realistic sensing scenarios \citep{diehl2008quantum,PhysRevA.49.4110}.

\section{ACKNOWLEDGMENTS}

The computational work in this study, from mathematical derivations
to visualizations, was carried out using version 14.3 of the Wolfram
Language and the Wolfram Quantum Framework. This work was supported
by the National Natural Science Foundation of China (Grant No. 12365005).

\appendix

\section{\label{sec:A1}Derivation of the Cramér-Rao bounds for the NOON,
bat, and entangled coherent states}

In this appendix, we derive the Cramér--Rao bounds for the NOON,
bat, and entangled coherent states discussed in the main text. The
phase to be estimated is encoded in mode b via the unitary transformation
$U(\phi)=e^{i\phi n_{b}}$. For a pure probe stat $|\psi(\phi)\rangle$,
the quantum Fisher information is $F_{Q}=4(\langle\partial_{\phi}\psi|\partial_{\phi}\psi\rangle-|\langle\psi|\partial_{\phi}\psi\rangle|^{2})$,
and the corresponding phase sensitivity obeys the quantum Cramér--Rao
bound $\delta\varphi\ge1/\sqrt{F_{Q}}$. In the present phase-encoding
scheme, the QFI reduces to four times the variance of $n_{b}=b^{\dagger}b$,
namely, $F_{Q}=4\,\mathrm{Var}(n_{b})=4(\langle n^{2}_{b}\rangle-\langle n_{b}\rangle^{2})$.

\subsection{Complete derivation for the NOON state}

The NOON state is $|{\rm NOON}\rangle=(|N,0\rangle+|0,N\rangle)/\sqrt{2}$.
Since mode $a$ and $b$ contains either $0$ or $N$ photons with
equal probability, we have $\langle n_{b}\rangle=N/2$ and $\langle n^{2}_{b}\rangle=N^{2}/2$.
Hence 
\begin{equation}
\mathrm{Var}(n_{a})=\langle n^{2}_{b}\rangle-\langle n_{b}\rangle^{2}=\frac{N^{2}}{2}-\frac{N^{2}}{4}=\frac{N^{2}}{4}.
\end{equation}

The quantum Fisher information is therefore $F^{{\rm N}}_{Q}=4\,\mathrm{Var}(n_{b})=N^{2}$,
and the QCRB becomes 
\begin{equation}
\delta\varphi_{N}=\frac{1}{N}.
\end{equation}

\subsection{Complete derivation for the bat state}

For even $N$, write $N=2j$. The bat states can be represented by
a two-mode Dickie as 
\begin{equation}
|B_{N}\rangle=\sum^{N/2}_{k=0}\frac{\sqrt{(N-2k)!\,(2k)!}}{k!\,(N/2-k)!\,\sqrt{2^{N}}}\,|N-2k\rangle_{1}|2k\rangle_{2}.
\end{equation}

This state is equivalently the $\pi/2$-rotation of the Dicke state
$|j,0\rangle$ in the Schwinger representation, $|B_{N}\rangle=e^{-i(\pi/2)J_{y}}|j,0\rangle$,
with $J_{z}=(n_{1}-n_{2})/2$, and $n_{b}=j-J_{z}$. Because $n_{a}=j-J_{z}$,
the variance of $n_{b}$ is the same as the variance of $J_{z}$:
\begin{equation}
\mathrm{Var}(n_{a})=\mathrm{Var}(J_{z}).
\end{equation}

We now evaluate the moments in the rotated state. First, $\langle J_{z}\rangle_{B}=\langle j,0|e^{i(\pi/2)J_{y}}J_{z}e^{-i(\pi/2)J_{y}}|j,0\rangle=\langle j,0|J_{x}|j,0\rangle=0$.
Next, $\langle J^{2}_{z}\rangle_{B}=\langle j,0|J^{2}_{x}|j,0\rangle$.
Using $J_{x}=(J_{+}+J_{-})/2$, we have $J^{2}_{x}=(J_{+}J_{-}+J_{-}J_{+}+J^{2}_{+}+J^{2}_{-})/4$.
When sandwiched between $\langle j,0|$ and $|j,0\rangle$, the terms
$J^{2}_{+}$ and $J^{2}_{-}$ do not contribute, so $\langle j,0|J^{2}_{x}|j,0\rangle=\langle j,0|J_{+}J_{-}+J_{-}J_{+}|j,0\rangle/4$.
Using the identity $J_{+}J_{-}+J_{-}J_{+}=2(J^{2}-J^{2}_{z})$, we
obtain $\langle J^{2}_{z}\rangle_{B}=\langle j,0|J^{2}-J^{2}_{z}|j,0\rangle/2$.
Since $J^{2}|j,0\rangle=j(j+1)|j,0\rangle$ and $J_{z}|j,0\rangle=0$,
this gives $\langle J^{2}_{z}\rangle_{B}=j(j+1)/2$. Therefore $\mathrm{Var}(n_{2})=\mathrm{Var}(J_{z})=j(j+1)/2$.
Substituting $j=N/2$, we find 
\begin{equation}
\mathrm{Var}(n_{b})=\frac{N(N+2)}{8}.
\end{equation}
Hence the quantum Fisher information is $F^{{\rm B}}_{Q}=4\,\mathrm{Var}(n_{b})=N(N+2)/2=N[(N/2)+1]$,
and the QCRB becomes 
\begin{equation}
\delta\varphi_{B}=\frac{1}{\sqrt{N[(N/2)+1]}}.
\end{equation}

\subsection{Complete derivation for the ECS}

The ECS state is $|{\rm ECS}\rangle=\mathcal{N}(|\alpha\rangle_{a}|0\rangle_{b}+|0\rangle_{a}|\alpha\rangle_{b})$,
with normalization $\mathcal{N}$. The mean photon number in mode
$b$ is 
\begin{align}
\langle n_{a}\rangle & =\mathcal{N}^{2}\left[\langle\alpha,0|n_{b}|\alpha,0\rangle+\langle0,\alpha|n_{b}|0,\alpha\rangle\right]\nonumber \\
 & =\mathcal{N}^{2}|\alpha|^{2}.
\end{align}
The cross terms vanish because $n|0\rangle=0$ and $\langle0|n=0$.
Similarly, using the coherent-state moments $\langle\alpha|n_{b}|\alpha\rangle=|\alpha|^{2}$,
and $\langle\alpha|n^{2}_{b}|\alpha\rangle=|\alpha|^{4}+|\alpha|^{2}$,
we obtain 
\begin{align}
\langle n^{2}_{a}\rangle & =\mathcal{N}^{2}\left[\langle\alpha,0|n^{2}_{b}|\alpha,0\rangle+\langle0,\alpha|n^{2}_{b}|0,\alpha\rangle\right]\nonumber \\
 & =\mathcal{N}^{2}(|\alpha|^{4}+|\alpha|^{2}).
\end{align}

Therefore, 
\begin{align}
\mathrm{Var}(n_{a}) & =\langle n^{2}_{b}\rangle-\langle n_{b}\rangle^{2}\nonumber \\
 & =\mathcal{N}^{2}|\alpha|^{2}\left[1+|\alpha|^{2}(1-\mathcal{N}^{2})\right].
\end{align}

The quantum Fisher information is thus 
\begin{equation}
F^{{\rm C}}_{Q}=4\mathcal{N}^{2}|\alpha|^{2}\left[1+|\alpha|^{2}\left(1-\mathcal{N}^{2}\right)\right],
\end{equation}
and the QCRB is 
\begin{equation}
\delta\varphi_{C}=\frac{1}{2\mathcal{N}|\alpha|\sqrt{1+|\alpha|^{2}(1-\mathcal{N}^{2})}}.
\end{equation}

\section{\label{sec:A2}Model and dynamical evolution}

The system is initialized in the general superposition state $|\psi_{i}\rangle=\cos(\theta/2)|0\rangle+e^{i\delta}\sin(\theta/2)|1\rangle$.
The pointer is prepared in a coherent-state superposition of the form
$|\phi\rangle$, the total initial state is therefore $|\Psi_{{\rm in}}\rangle=|\psi_{i}\rangle\otimes|\phi\rangle$.
The interaction between the system and the pointer is described by
$\hat{H}_{\mathrm{int}}=g_{a}\hat{\sigma}_{x}\otimes\hat{P}_{x}+g_{b}\hat{\sigma}_{y}\otimes\hat{P}_{y}$.
The corresponding unitary evolution is $U(t)=e^{-iH_{{\rm int}}}$.
After the interaction, the composite state becomes $|\Psi_{{\rm after}}\rangle=U(t)|\Psi_{{\rm in}}\rangle.$
A projective measurement is performed on the system, postselecting
onto a target state $|\psi_{f}\rangle$. The projector is 
\begin{equation}
\Pi_{{\rm post}}=|\psi_{f}\rangle\langle\psi_{f}|\otimes I.
\end{equation}

The unnormalized post-selected state is $|\widetilde{\Psi}_{{\rm post}}\rangle=\Pi_{{\rm post}}|\Psi_{{\rm after}}\rangle$.
The success probability reads $P_{{\rm s}}=\langle\Psi_{{\rm after}}|\Pi_{{\rm post}}|\Psi_{{\rm after}}\rangle$.
The normalized c-selected state is $|\Psi_{{\rm post}}\rangle=|\widetilde{\Psi}_{{\rm post}}\rangle/\sqrt{P_{{\rm s}}}$.
After post-selection, the system undergoes dissipative evolution described
by the Lindblad master equation. The dissipation channels are modeled
by the Lindblad operators $L_{a}=\sqrt{\gamma_{a}}a$, and $L_{b}=\sqrt{\gamma_{b}}b$,
where $\gamma_{a}$ and $\gamma_{b}$ are decay rates. The density
operator $\rho(\tau)$ satisfies 
\begin{equation}
\frac{d\rho}{d\tau}=-i[H_{{\rm int}},\rho]+\sum_{j=a,b}(L_{j}\rho L^{\dagger}_{j}-\{L^{\dagger}_{j}L_{j},\rho\}/2),
\end{equation}
with initial condition $\rho(0)=|\Psi_{{\rm post}}\rangle\langle\Psi_{{\rm post}}|.$
To evaluate the entanglement between the two pointer modes, we trace
out the system degrees of freedom, $\rho_{{\rm ptr}}(\tau)=\mathrm{Tr}_{{\rm s}}[\rho(\tau)]$.
We then compute the entanglement negativity via partial transposition,
\begin{equation}
N_{e}(\tau)=\frac{\left\Vert \rho^{T_{a}}_{{\rm ptr}}(\tau)\right\Vert _{1}-1}{2},
\end{equation}
where $T_{a}$ denotes partial transpose with respect to subsystem
$a$.

\bibliographystyle{apsrev4-2}
\bibliography{ref-ECS}

\begin{thebibliography}{62}
\bibitem [{eq()}]{eq}%
  \BibitemOpen
  \href@noop {} {}\bibinfo {note} {$|\psi_N\rangle = (|N\rangle_a|0\rangle_b + |0\rangle_a|N\rangle_b)/\sqrt{2}$ and $|\psi_B\rangle = \sum_{k=0}^{N/2} \sqrt{(N-2k)!}\sqrt{(2k)!}/k!(N/2-k)!\sqrt{2^N} |N-2k\rangle_a|2k\rangle_b$. The bat state is so named because its amplitude distribution resembles the wings of a bat.}\BibitemShut {Stop}%
\end{thebibliography}

\end{document}